\documentclass[aps,prd,twocolumn,superscriptaddress,nofootinbib,longbibliography,floatfix]{revtex4-2}

\usepackage{amsmath,amssymb,amsfonts}
\usepackage{bm}
\usepackage{booktabs}
\usepackage{graphicx}
\usepackage{microtype}
\usepackage[colorlinks=true,linkcolor=blue,citecolor=blue,urlcolor=blue]{hyperref}

\newcommand{\hC}{\widehat{C}}
\newcommand{\hCD}{\widehat{C}_{D}}
\newcommand{\hH}{\widehat{H}}
\newcommand{\Ai}{\operatorname{Ai}}
\newcommand{\Bi}{\operatorname{Bi}}
\newcommand{\dd}{\mathrm{d}}
\newcommand{\PV}{\operatorname{PV}}
\newcommand{\sgn}{\operatorname{sgn}}
\newcommand{\GR}{G_{\mathbb{R}}}
\newcommand{\GRD}{G_{\mathbb{R},D}}
\newcommand{\Gp}{G_{+}}
\newcommand{\Gm}{G_{-}}
\newcommand{\GpD}{G_{+,D}}
\newcommand{\Rb}{\mathbb{R}}
\newcommand{\Nt}{\widetilde{N}}
\newcommand{\eps}{\epsilon}

\begin{document}

\title{Exact DeWitt kernels in Lorentzian quantum cosmology}

\author{Hiroki Matsui}
\email{hiroki.matsui@yukawa.kyoto-u.ac.jp}
\affiliation{Osaka Central Advanced Mathematical Institute (OCAMI), Osaka Metropolitan
University, 3-3-138 Sugimoto, Sumiyoshi, Osaka 558-8585, Japan}
\affiliation{TRIP Headquarters, RIKEN, Wako 351-0198, Japan}

\begin{abstract}
The DeWitt boundary condition can be implemented in the Lorentzian path integral of minisuperspace quantum cosmology by summing only over non-singular paths. In this paper, we present a general definition of the resulting DeWitt kernels, along with specific rules for their construction, and evaluate them rigorously. The lapse integral over the positive half-line yields the Dirichlet Green function of the Wheeler--DeWitt operator, whereas the integral over the whole real line yields the Dirichlet group-averaging kernel; both vanish when either endpoint lies at the singularity. Since the minisuperspace actions are at most quadratic, all lapse integrals are performed in closed form without saddle-point approximations. We obtain the kernels for the closed universe without cosmological constant, for the flat de Sitter universe, and, in terms of Airy functions, for the closed and open de Sitter universes. Throughout, the DeWitt condition is imposed as the boundary condition at the singularity, which selects a self-adjoint constraint operator on the half-line and fixes the kernels uniquely, including their normalization.
\end{abstract}

\maketitle

\section{Introduction}
\label{sec:intro}
The DeWitt boundary condition~\cite{DeWitt:1967yk} requires the wave function of the universe to vanish on singular three-geometries. It has long been discussed as a mechanism for avoiding the big-bang singularity in canonical quantum gravity~\cite{Kiefer:2004xyv,Hajicek:2001yd,Kiefer:2019bxk,Matsui:2021yte,Martens:2022dtd}, where it is imposed as a boundary condition on solutions of the Wheeler--DeWitt equation. The other classic proposals for the quantum state of the universe, the no-boundary~\cite{Hartle:1983ai} and tunneling~\cite{Vilenkin:1984wp,Vilenkin:1986cy} proposals, are formulated most naturally as gravitational path integrals, and their Lorentzian path-integral formulation has been studied intensively in recent years~\cite{Feldbrugge:2017kzv,Feldbrugge:2017mbc,Lehners:2023yrj,Honda:2024aro}. It is therefore natural to ask how the DeWitt condition is expressed in the same language.

In Ref.~\cite{Matsui:2023tkw} it was proposed to implement the DeWitt condition by summing only over non-singular paths. For the squared scale factor $q=a(t)^{2}$, the lapse $N$ and the Hamiltonian constraint $C(q,p)$, the resulting DeWitt kernel is
\begin{equation}
G_{\rm DW}[q_1;q_0]=\int_{\mathcal C}\!\dd N\!\int_{q(t)>0}\!\mathcal{D}q\,\mathcal{D}p\,e^{\,iS[q,p,N]},
\label{eq:GDW_intro}
\end{equation}
with $S[q,p,N]=\int_{0}^{1}\dd t\,(p\dot q-NC)$ and $q(0)=q_0$, $q(1)=q_1$, where $\mathcal C$ is the range of the lapse in the Batalin--Fradkin--Vilkovisky (BFV) path integral~\cite{Fradkin:1975cq,Batalin:1977pb,Halliwell:1988wc}. When the constraint is reflection symmetric about $q=0$, the restriction to $q(t)>0$ is implemented by the method of images~\cite{Janke:1979fv,Goodman:1981,Auerbach:1997,Farhi:1989jz},
\begin{equation}
G_{\rm DW}[q_1;q_0]=G[q_1;q_0]-G[q_1;-q_0],
\label{eq:image_intro}
\end{equation}
where $G$ is the unrestricted path integral over $-\infty<q(t)<\infty$: paths that cross $q=0$ are cancelled pairwise by their reflected images. Jia~\cite{Jia:2024comment} emphasized that such a kernel must vanish when \emph{either} endpoint lies at the singularity,
\begin{equation}
G_{\rm DW}[0;q_0]=0=G_{\rm DW}[q_1;0],
\label{eq:twosided}
\end{equation}
since no path confined to $q>0$ can start or end at $q=0$. Jia also proposed a closed-form Dirichlet kernel for non-symmetric potentials~\cite{Jia:2024comment}; see also~\cite{Jia:2022gzo,Jia:2022nda}.

In this paper we construct DeWitt kernels exactly, and in a form that can be applied directly to other minisuperspace models. We find that the results crucially depend on two key factors: the lapse contour and the boundary reflection. The former involves the integration range of the lapse. The integral over $\mathcal C=(0,\infty)$ yields the Green function $-i(\hC-i0)^{-1}$ of the Wheeler--DeWitt operator $\hC$, which obeys an inhomogeneous Wheeler--DeWitt equation and has the causal interpretation of Teitelboim~\cite{Teitelboim:1981ua,Teitelboim:1983fh}; the integral over $\mathcal C=\Rb$ yields the group-averaging kernel $2\pi\delta(\hC)$ of refined algebraic quantization~\cite{Ashtekar:1995zh,Marolf:2000iq}, which obeys the homogeneous equation~\cite{Halliwell:1988wc}. The DeWitt condition can be imposed on either, but the resulting kernels are different objects and must not be confused. As for the latter, since lapse-integrated kernels typically depend on $|q_1-q_0|$, the image term in Eq.~\eqref{eq:image_intro} must be evaluated with a kernel valid for all orderings of the endpoints. Throughout, the DeWitt condition is imposed as the Dirichlet self-adjoint extension of the half-line constraint operator, which fixes the kernels, including their normalization, through operator identities.

Because the minisuperspace actions considered here are at most quadratic in the configuration variable, the fixed-lapse kernel is an exact Gaussian and every lapse integral can be performed in closed form, without saddle-point or Picard--Lefschetz approximations~\cite{Witten:2010cx,Feldbrugge:2017kzv,Halliwell:1988ik,Honda:2024aro}. For the closed universe with $\Lambda=0$ we find the positive-lapse DeWitt kernel $i\,e^{-q_>/2}\sinh(q_</2)$, while the whole-real-line kernel vanishes identically because zero lies outside the spectrum of the constraint. For the flat universe with $\Lambda>0$, in the variable $x=\tfrac23q^{3/2}$, the whole-real-line kernel is $(2/\sqrt{\lambda})\sin(kx_1)\sin(kx_0)$ with $k=\sqrt{\lambda}/2$, whose nodes lie at $L_{n}=(2nG/H_{\Lambda})^{1/3}$. For the linear potential of the closed and open de Sitter models, where the method of images cannot be applied to the known full-line kernel, we obtain exact Airy-function kernels by a rank-one formula, which we derive as an infinite-barrier limit and show to be equivalent to the method of images for the even extension of the potential. Whenever it is non-vanishing, the whole-real-line DeWitt kernel factorizes into a product of the DeWitt solution at the two endpoints.

The paper is organized as follows. In Sec.~\ref{sec:model} we introduce the minisuperspace model and the three kernels obtained from the BFV path integral with a fixed lapse, a positive lapse, and a lapse over the whole real line. In Sec.~\ref{sec:def} we give the general definition of DeWitt kernels and collect three rules for constructing them, and in Sec.~\ref{sec:images} we justify the method of images. In Sec.~\ref{sec:exact} we compute the exact fixed-lapse kernel and evaluate the DeWitt kernels for constant potentials and for the linear potential; the results are summarized in Table~\ref{tab:summary}. Section~\ref{sec:discussion} is devoted to discussion. Appendix~\ref{app:integral} contains the lapse integral used in the text, and Appendix~\ref{app:rankone} the derivation of the rank-one formula.

Throughout we set $\hbar=1$, normalize $\langle q|q'\rangle=\delta(q-q')$ and $\langle q|p\rangle=e^{ipq}/\sqrt{2\pi}$, and denote the spatial curvature by $\kappa\in\{-1,0,1\}$.

\section{Minisuperspace model and lapse integrals}
\label{sec:model}

\subsection{Constraint operator and BFV path integral}
\label{sec:constraint}
We consider the homogeneous and isotropic metric~\cite{Halliwell:1988wc,Feldbrugge:2017kzv,Matsui:2023tkw}
\begin{equation}
\dd s^{2}=\sigma^{2}\left[-\frac{N^{2}}{q}\,\dd t^{2}+q\,\dd\Omega_{\kappa}^{2}\right],\qquad \sigma^{2}=\frac{2G}{3\pi},
\label{eq:metric}
\end{equation}
with $q=a^{2}$ and the rescaled cosmological constant $\lambda=\sigma^{2}\Lambda/3$. With the spatial volume normalized as in Refs.~\cite{Halliwell:1988wc,Feldbrugge:2017kzv}, the Einstein--Hilbert action with the Gibbons--Hawking--York term reduces to
\begin{equation}
S[q,N]=\frac12\int_{0}^{1}\dd t\,N\left[-\frac{\dot q^{2}}{4N^{2}}+\kappa-\lambda q\right].
\label{eq:action}
\end{equation}
The canonical momentum is $p=\partial L/\partial\dot q=-\dot q/(4N)$, so that $\dot q=-4Np$, and the Legendre transformation gives $H_{\rm tot}=NC$ with
\begin{equation}
C(q,p)=-2p^{2}+\frac{\lambda q-\kappa}{2}.
\label{eq:constraint}
\end{equation}
We quantize with the flat measure $\dd q$ and $p=-i\,\dd/\dd q$, which yields the constraint operator and the Wheeler--DeWitt equation
\begin{equation}
\hC=2\frac{\dd^{2}}{\dd q^{2}}+\frac{\lambda q-\kappa}{2},\qquad
\left[4\frac{\dd^{2}}{\dd q^{2}}+\lambda q-\kappa\right]\Psi(q)=0.
\label{eq:WDW}
\end{equation}
Other operator orderings and other choices of the measure define different quantum theories~\cite{Hawking:1985bk,Kontoleon:1998pw,Mondal:2025qyd}; we do not pursue them here, but we shall see below that the change of variables commonly used for the flat universe amounts to such a change.

In the BFV path integral with the gauge $\dot N=0$, the integration over the multiplier $\Pi$ conjugate to $N$ restricts $N$ to a constant, the ghost integrals cancel against the measure, and one is left with an ordinary integral over $N$ of the fixed-lapse kernel~\cite{Halliwell:1988wc},
\begin{equation}
G[q_1;q_0]=\int_{\mathcal C}\dd N\,K_{N}(q_1,q_0),\quad
K_{N}=\langle q_1|e^{-iN\hC}|q_0\rangle ,
\label{eq:BFV}
\end{equation}
where $\mathcal C$ is the range (or contour) of the lapse. In the phase-space discretization the kernel carries one momentum integral $\dd p/(2\pi)$ per time step, the factor $1/(2\pi)$ being required by $K_{0}=\delta(q_1-q_0)$; it fixes the overall normalization of all kernels below.

\subsection{Fixed lapse, positive lapse, and whole real line}
\label{sec:three}
The fixed-lapse kernel satisfies
\begin{equation}
i\,\partial_{N}K_{N}(q_1,q_0)=\hC_{q_1}K_{N}(q_1,q_0),\quad K_{0}=\delta(q_1-q_0),
\label{eq:KN_eq}
\end{equation}
and is, for $N\neq0$, \emph{not} a solution of the Wheeler--DeWitt equation. Because $\hC$ is a real differential operator its kernel is symmetric, $K_{N}(q_1,q_0)=K_{N}(q_0,q_1)$, and
\begin{equation}
K_{-N}(q_1,q_0)=K_{N}(q_1,q_0)^{*}.
\label{eq:KN_conj}
\end{equation}

Integrating over positive lapse requires a convergence factor. For $\eps>0$,
\begin{equation}
\int_{0}^{\infty}\dd N\,e^{-\eps N}e^{-iN\hC}=(\eps+i\hC)^{-1}=-i(\hC-i\eps)^{-1},
\end{equation}
so that the positive-lapse integral is given by the boundary value of the resolvent at the origin, approached from the upper half-plane,
\begin{align}
\Gp&:=\lim_{\eps\downarrow0}\int_{0}^{\infty}\dd N\,e^{-\eps N}K_{N}\nonumber\\
&=-i(\hC-i0)^{-1}=\pi\delta(\hC)-i\,\PV\frac{1}{\hC},
\label{eq:Gplus_def}
\end{align}
where the last form follows from the Sokhotski--Plemelj formula applied to the spectral representation of $\hC$. It obeys the inhomogeneous Wheeler--DeWitt equation
\begin{equation}
\hC_{q_1}\Gp(q_1,q_0)=-i\,\delta(q_1-q_0),
\label{eq:Gplus_eq}
\end{equation}
which may also be derived by integrating Eq.~\eqref{eq:KN_eq} by parts~\cite{Feldbrugge:2017kzv,Teitelboim:1981ua}. The factor $e^{-\eps N}$ is not a new physical ingredient: it is the Abel regulator that specifies from which side the spectral point $\hC=0$ is approached, and thereby selects the corresponding Green function.\footnote{For a relativistic particle, taking $\hC=-(p^{2}-m^{2})$ and $K_{N}=e^{-iN\hC}$, Eq.~\eqref{eq:Gplus_def} gives
$\Gp=\frac{i}{p^{2}-m^{2}+i0}$, 
which is the standard Feynman propagator~\cite{Teitelboim:1981ua}. Thus, with the conventions used here, the positive-lapse integral reproduces the usual Feynman $i0$ prescription. The apparent sign of the $i0$ term depends on the convention for $\hC$, the phase in $K_N$, and the choice of lapse half-line.}
Integrating over the negative half-line instead gives $\Gm=i(\hC+i0)^{-1}=\pi\delta(\hC)+i\,\PV\hC^{-1}$, and by Eq.~\eqref{eq:KN_conj}
\begin{equation}
\Gm(q_1,q_0)=\Gp(q_1,q_0)^{*}.
\label{eq:Gminus}
\end{equation}

Integrating over the whole real line, with the symmetric regulator $e^{-\eps|N|}$, the principal-value parts cancel and
\begin{equation}
\GR:=\lim_{\eps\downarrow0}\int_{-\infty}^{\infty}\dd N\,e^{-\eps|N|}K_{N}=2\pi\delta(\hC)=2\,\mathrm{Re}\,\Gp .
\label{eq:GR_def}
\end{equation}
The whole-real-line kernel is $\GR=2\pi\delta(\hC)$ and therefore satisfies the Wheeler--DeWitt equation in both arguments,
\begin{equation}
\hC_{q_1}\GR=\hC_{q_0}\GR=0,
\end{equation}
as follows directly from the spectral identity $\hC\delta(\hC)=0$. It is the group-averaging kernel that defines the rigging map in refined algebraic quantization~\cite{Ashtekar:1995zh,Marolf:2000iq,Ashtekar:2010ve}.

We will repeatedly use the following properties of $\GR=2\pi\delta(\hC)$.  
(i) $\GR$ is real and symmetric, and defines a positive semidefinite quadratic form,
\begin{equation}
\int \dd q_1\dd q_0\,\phi(q_1)^*\GR(q_1,q_0)\phi(q_0)\geq0,
\end{equation}
as follows from the positivity of the spectral measure $\delta(\hC)$.  
(ii) If zero lies outside the spectrum of $\hC$, then $\delta(\hC)=0$ and hence $\GR$ vanishes, whereas the positive-lapse kernel reduces to the ordinary resolvent $\Gp=-i\hC^{-1}$.  
(iii) Although $\delta(\hC)$ formally restricts the kernel to the constraint surface $\hC=0$, it is not an ordinary projection operator: its square contains the divergent gauge-volume factor $\delta(0)$. This is the familiar group-averaging structure underlying the rigging map in refined algebraic quantization~\cite{Marolf:2000iq}.

The regulators leave the integrand unchanged at $N=0$. This causes no difficulty, since $K_N\to\delta(q_1-q_0)$ as $N\to0$ and the $|N|^{-1/2}$ behavior of the explicit kernels below is locally integrable. The whole-real-line integral can therefore be decomposed into its positive- and negative-lapse parts~\cite{Jia:2024comment}. The choice of lapse range determines the resulting object: integration over $N>0$ gives the causal Green function $\Gp$~\cite{Teitelboim:1981ua,Teitelboim:1983fh}, whereas integration over $N\in\mathbb R$ gives the group-averaging kernel $\GR=2\pi\delta(\hC)$ satisfying the homogeneous Wheeler--DeWitt equation~\cite{Halliwell:1988wc}. This correspondence applies here to a single self-adjoint constraint with the flat measure $\dd N$. In the full gravitational path integral, gauge fixing, the Faddeev--Popov measure, and the treatment of the lapse contour near $N=0$ introduce additional subtleties~\cite{Banihashemi:2024aal}.

\section{Definition and construction of DeWitt kernels}
\label{sec:def}
In this section we state the definition of DeWitt kernels and the rules for constructing them in a form independent of the particular model. The only input is a constraint operator of the form $\hC=2\,\dd^{2}/\dd q^{2}+U(q)$ with real $U$, as in Eq.~\eqref{eq:WDW}, and the normalization of Eq.~\eqref{eq:BFV}.

\subsection{Path-integral definition}
\label{sec:def_path}
For $q_0,q_1>0$ let
\begin{equation}
K_{N,D}(q_1,q_0)=\int_{q(t)>0}\mathcal{D}q\,\mathcal{D}p\;\exp\left[i\int_{0}^{1}\dd t\,(p\dot q-NC)\right]
\label{eq:KND_path}
\end{equation}
be the fixed-lapse path integral over non-singular paths with $q(0)=q_0$ and $q(1)=q_1$. The DeWitt kernel with lapse range $\mathcal C$ is
\begin{equation}
G^{(\mathcal C)}_{\rm DW}(q_1,q_0)=\int_{\mathcal C}\dd N\,K_{N,D}(q_1,q_0),
\label{eq:GDW_def}
\end{equation}
which is Eq.~\eqref{eq:GDW_intro}. With the regulators of Eqs.~\eqref{eq:Gplus_def} and \eqref{eq:GR_def} understood, the two lapse ranges of interest define
\begin{equation}
\GpD\equiv G^{(0,\infty)}_{\rm DW},\qquad \GRD\equiv G^{(\Rb)}_{\rm DW}=2\,\mathrm{Re}\,\GpD ,
\label{eq:GDW_pm}
\end{equation}
where the last equality follows from $K_{-N,D}=K_{N,D}^{*}$ as in Eq.~\eqref{eq:Gminus}.

\subsection{Operator definition}
\label{sec:def_op}

To define the half-line path integral in Eq.~\eqref{eq:KND_path},
we must specify the constraint operator $\hC=2\,\dd^{2}/\dd q^{2}+U(q)$ 
on $L^{2}((0,\infty),\dd q)$, together with a boundary condition
at $q=0$. This boundary condition must be chosen so that $\hC$
is self-adjoint. For the real potentials considered here,
$q=0$ is a regular endpoint, and no additional boundary condition
is required at $q\to\infty$. Integration by parts gives
\begin{equation}
\langle\phi,\hC\psi\rangle-\langle\hC\phi,\psi\rangle
=
-2\bigl[\phi^{*}\psi'-\phi'^{*}\psi\bigr]_{q=0},
\label{eq:boundaryform}
\end{equation}
where the contribution at infinity vanishes for functions in
the operator domain. The self-adjoint extensions are specified
by the one-parameter family of boundary conditions
\begin{equation}
\cos\alpha\,\Psi(0)+\sin\alpha\,\Psi'(0)=0,
\qquad
\alpha\in[0,\pi),
\label{eq:robin}
\end{equation}
with the same value of $\alpha$ for all functions in the
domain~\cite{Bonneau:1999zq,Farhi:1989jz,Carreau:1990wh,daLuz:1995}.
\footnote{More precisely, $q\to\infty$ is a limit-point endpoint,
and the minimal symmetric operator has deficiency indices $(1,1)$.
Self-adjointness requires the operator and its adjoint to have
the same domain; making the boundary term vanish on an arbitrarily
restricted domain is not sufficient.}
For $\alpha\ne0$, Eq.~\eqref{eq:robin} can be written as the
Robin condition $\Psi'(0)=\eta\,\Psi(0)$, with
$\eta=-\cot\alpha\in\mathbb{R}$; $\alpha=\pi/2$ gives the Neumann
condition $\Psi'(0)=0$. The DeWitt condition $\Psi(0)=0$ selects
the Dirichlet extension, $\alpha=0$. We denote the resulting
self-adjoint constraint operator by $\hCD$.

Since $\hCD$ is self-adjoint, $e^{-iN\hCD}$ generates unitary
evolution in the lapse parameter $N$. The corresponding
fixed-lapse kernel is
\begin{equation}
K_{N,D}(q_1,q_0)
=
\langle q_1|e^{-iN\hCD}|q_0\rangle.
\end{equation}
For the real potential and Dirichlet boundary condition
considered here, this kernel satisfies
$K_{-N,D}(q_1,q_0)=K_{N,D}(q_1,q_0)^*$. 
Using the regulated lapse integrals of Sec.~\ref{sec:three},
we obtain the positive-lapse Green function and the
whole-real-line group-averaging kernel,
\begin{equation}
\GpD=-i(\hCD-i0)^{-1},
\qquad
\GRD=2\pi\delta(\hCD).
\label{eq:GDW_operator}
\end{equation}
For $\eps>0$, the resolvent $(\hCD-i\eps)^{-1}$ is well defined
because the spectrum of $\hCD$ is real. For the models considered
below, the regulator limits are understood in the distributional
sense where necessary. The spectral representations of these
kernels use the $\delta$-normalized generalized eigenfunctions
introduced below.

Different self-adjoint extensions impose different boundary
conditions at $q=0$ and generally lead to different kernels
and spectra. In the path-integral description, this freedom
is reflected in the weights assigned to paths that reach
the boundary~\cite{Farhi:1989jz,Carreau:1990wh}.
The DeWitt condition $\Psi(0)=0$ selects the Dirichlet
extension, $\alpha=0$, and all kernels below refer to
this choice.

\subsection{Defining properties}
\label{sec:def_prop}

The DeWitt kernels defined in Eq.~\eqref{eq:GDW_operator} are symmetric under $q_1\leftrightarrow q_0$ and satisfy the Dirichlet condition at both endpoints, as expressed in Eq.~\eqref{eq:twosided}. They also obey the operator identities
\begin{equation}
\hat C_D\,G_{+,D}=-i\,\mathbf 1,
\qquad
\hat C_D\,G_{\mathbb R,D}=0.
\label{eq:D_identities}
\end{equation}
These properties provide a direct criterion for identifying a DeWitt kernel. In particular, imposing the boundary condition at only one endpoint is insufficient: the kernel must vanish when either $q_0=0$ or $q_1=0$. Any candidate DeWitt kernel should therefore be checked both against the two-sided boundary condition \eqref{eq:twosided} and against the operator identities \eqref{eq:D_identities}.

\subsection{Construction rules}
\label{sec:rules}
The DeWitt kernels can be obtained from the unrestricted full-line kernels $K_{N}$, $\Gp$ and $\GR$ of Eqs.~\eqref{eq:BFV}, \eqref{eq:Gplus_def} and \eqref{eq:GR_def} in three equivalent ways.

(a) \emph{Method of images.} If $\hC$ commutes with the parity operator $(P\Psi)(q)=\Psi(-q)$, then
\begin{equation}
K_{N,D}(q_1,q_0)=K_{N}(q_1,q_0)-K_{N}(q_1,-q_0),
\label{eq:image_KN}
\end{equation}
and, since the lapse integral is linear,
\begin{align}
\GpD(q_1,q_0)&=\Gp(q_1,q_0)-\Gp(q_1,-q_0),\nonumber\\
\GRD(q_1,q_0)&=\GR(q_1,q_0)-\GR(q_1,-q_0),
\label{eq:image_G}
\end{align}
which is Eq.~\eqref{eq:image_intro}. The full-line kernels must be used in a form valid for all orderings of the endpoints; reflecting the final instead of the initial point gives the same result.

(b) \emph{Rank-one formula.} For an arbitrary real potential,
\begin{equation}
\GpD(q_1,q_0)=\Gp(q_1,q_0)-\frac{\Gp(q_1,0)\,\Gp(0,q_0)}{\Gp(0,0)},
\label{eq:rankone}
\end{equation}
provided $\Gp(0,0)\neq0$, and $\GRD=2\,\mathrm{Re}\,\GpD$. This formula, proposed in Ref.~\cite{Jia:2024comment}, is derived in Appendix~\ref{app:rankone} as the infinite-strength limit of a $\delta$-function barrier at $q=0$. This identity applies to the resolvent $\Gp$, but not to the fixed-lapse kernel $K_N$, and reduces to rule (a) when the method of images is applicable.

(c) \emph{Direct half-line construction.} 
Let $u_D$ be the solution of $\hC u=0$ satisfying the Dirichlet boundary condition $u_D(0)=0$. Let $u_R$ be the solution selected at $q\to\infty$ by the $-i0$ prescription, namely the $\eps\downarrow0$ limit of the solution of $(\hC-i\eps)u=0$, with $\eps>0$, that is square integrable as $q\to\infty$. The positive-lapse Dirichlet Green function is then
\begin{equation}
\GpD(q_1,q_0)
=
-\frac{i\,u_D(q_<)\,u_R(q_>)}{2\,W[u_D,u_R]},
\label{eq:GpD_direct}
\end{equation}
where $q_<\equiv\min(q_0,q_1)$, $q_>\equiv\max(q_0,q_1)$, and $W[f,g]=fg'-f'g$ is the Wronskian. The corresponding full-line Green function $\Gp$ is obtained by replacing $u_D$ with the solution selected by the same $-i0$ prescription at $q\to-\infty$, while $u_R$ remains the solution selected at $q\to\infty$.

Section~\ref{sec:images} justifies rule (a), and Sec.~\ref{sec:exact} applies these steps to three models.

\section{Method of images on the half-line}
\label{sec:images}
In this section we justify rule (a) of Sec.~\ref{sec:rules} and state explicitly the condition under which the method of images applies. Consider a Hamiltonian on the full line,
\begin{equation}
\hH=-\frac{1}{2m}\partial_x^2+V(x),
\end{equation}
where $V(x)$ is regular at the origin. Let $P$ denote the parity operator,
$(P\psi)(x)=\psi(-x)$.
If the potential is reflection symmetric, $V(-x)=V(x)$, then $[\hH,P]=0$. The full-line evolution kernel
\begin{equation}
K(x,y;T)=\langle x|e^{-i\hH T}|y\rangle
\end{equation}
therefore satisfies
\begin{align}
\begin{split}
K(-x,-y;T)&=K(x,y;T), \\
K(-x,y;T)&=K(x,-y;T).
\label{eq:Ksym}
\end{split}
\end{align}

Parity invariance does not require the wave function itself to be even or odd. Rather, it implies that the even and odd parity sectors evolve independently. The half-line Dirichlet problem is obtained by restricting the full-line theory to the odd sector~\cite{Dluhy:2011tbs}. To see this, let $\psi(y)$ be a wave function on $y>0$, and extend it to the full line as an odd function,
\begin{equation}
\psi_{\rm odd}(-y)=-\psi_{\rm odd}(y),
\end{equation}
with $\psi_{\rm odd}(y)=\psi(y)$ for $y>0$.
For $x>0$, its time evolution is
\begin{align}
\psi(x,T)
&=\int_{-\infty}^{\infty}\dd y\,K(x,y;T)\psi_{\rm odd}(y)
\nonumber\\
&=\int_{0}^{\infty}\dd y\,
\bigl[K(x,y;T)-K(x,-y;T)\bigr]\psi(y).
\end{align}
The corresponding half-line Dirichlet kernel is therefore
\begin{align}
K_D(x,y;T)
&=K(x,y;T)-K(x,-y;T)
\nonumber\\
&=K(x,y;T)-K(-x,y;T),
\label{eq:images}
\end{align}
where the second equality follows from Eq.~\eqref{eq:Ksym}.

Equation~\eqref{eq:images} immediately gives the Dirichlet boundary condition in both arguments,
\begin{equation}
K_D(0,y;T)=0,
\qquad
K_D(x,0;T)=0.
\end{equation}
It also has the correct initial condition. Indeed, for $x,y>0$,
\begin{equation}
K_D(x,y;0)
=
\delta(x-y)-\delta(x+y)
=
\delta(x-y),
\end{equation}
since $\delta(x+y)$ has no support in the interior of the half-line. Moreover, $K_D$ satisfies the Schr\"odinger equation in the final coordinate $x$, because each term in Eq.~\eqref{eq:images} does so.

If the full-line potential is not even, the two reflections are no longer equivalent and neither of them produces a Dirichlet kernel. Acting with the Schr\"odinger operator on the reflected kernel gives
\begin{equation}
\bigl(i\partial_{T}-\hH_{x}\bigr)K(-x,y;T)=\bigl[V(-x)-V(x)\bigr]K(-x,y;T),
\label{eq:asym}
\end{equation}
so the final-point reflection $K(x,y)-K(-x,y)$ vanishes at $x=0$ but does not solve the Schr\"odinger equation, while the initial-point reflection $K(x,y)-K(x,-y)$ solves the equation but does not vanish at $x=0$. Reflection symmetry is thus a necessary condition for the method of images. It must, however, be read as a condition on the auxiliary full-line problem, not on the physical half-line potential. Given any half-line potential $V_{+}(x)$, $x>0$, one may define the even extension $V_{\rm even}(x)=V_{+}(|x|)$, to which Eq.~\eqref{eq:images} applies exactly; the half-line Dirichlet problem is always the odd sector of \emph{some} reflection-symmetric problem. What fails for a non-symmetric potential such as $\lambda q-\kappa$ is only that the kernel of the even extension $\lambda|q|-\kappa$ is not the known Airy kernel of the linear potential, so that the method of images does not deliver a closed formula for free. Rules (b) and (c) circumvent this obstacle for the lapse-integrated kernels.

In cosmological applications, the operator that is reflection-symmetric is the full constraint operator, including its kinetic term, ordering, and inner product measure, as these determine $[\hC,P]=0$. For flat-measure quantization, this reduces to the evenness of the potential $\lambda q-\kappa$, which holds only for $\lambda=0$. For a flat universe, one may instead first perform a change of variables that renders the potential constant~\cite{Matsui:2023tkw}, the consequences of which we examine below.

Finally, the reflection must be applied to a kernel valid for all orderings of the endpoints~\cite{Jia:2024comment}. Lapse-integrated kernels typically depend on $|q_1-q_0|$, and a closed form derived under an ordering assumption, such as $q_0<q_1$, cannot be extended to a reflected point through substitution.

\section{Exact DeWitt kernels}
\label{sec:exact}

\subsection{Exact fixed-lapse kernel}
\label{sec:fixedlapse}
For the action \eqref{eq:action}, the phase-space path integral can be evaluated exactly. Integrating first over $q(t)$ imposes $\dot p=-N\lambda/2$,
so that the remaining functional integral reduces to an ordinary integral over a single momentum variable. With the correctly normalized measure one
finds
\begin{equation}
K_{N}(q_1,q_0)
=
\int_{-\infty}^{\infty}\frac{\dd p}{2\pi}\,
e^{iF(N,p)},
\label{eq:KN_p}
\end{equation}
where
\begin{equation}
F(N,p)
=
2Np^{2}
-\lambda N^{2}p
+\frac{\lambda^{2}N^{3}}{6}
+\frac{\kappa-\lambda q_1}{2}N
+p(q_1-q_0),
\label{eq:F}
\end{equation}
in agreement with Ref.~\cite{Halliwell:1988wc}. The remaining momentum integral is Gaussian. For $N>0$, it gives
\begin{equation}
K_{N}(q_1,q_0)
=
\sqrt{\frac{i}{8\pi N}}\,
e^{iS_{\rm cl}(q_1,q_0;N)},
\label{eq:KN_exact}
\end{equation}
with
\begin{equation}
S_{\rm cl}(q_1,q_0;N)
=
\frac{\lambda^{2}N^{3}}{24}
-\frac{N}{4}\bigl[\lambda(q_1+q_0)-2\kappa\bigr]
-\frac{(q_1-q_0)^{2}}{8N},
\label{eq:Scl}
\end{equation}
in agreement with Refs.~\cite{Halliwell:1988ik,Feldbrugge:2017kzv}. Since the action is quadratic in the dynamical variable, Eq.~\eqref{eq:KN_exact} is the exact fixed-lapse kernel rather than a semiclassical approximation.
For $N<0$, the phase of the square root is fixed by the relation \eqref{eq:KN_conj}. The result for arbitrary nonzero lapse can therefore be written as
\begin{equation}
K_{N}(q_1,q_0)
=
\frac{e^{i\pi\,\sgn(N)/4}}{\sqrt{8\pi|N|}}\,
e^{iS_{\rm cl}(q_1,q_0;N)},
\qquad
N\neq0, 
\label{eq:KN_allN}
\end{equation}
and the phase $e^{i\pi/4}$ for $N>0$ is the one compatible with $K_{N}\to\delta(q_1-q_0)$ as $N\downarrow0$.

As a consistency check on both the functional form and the normalization, we next evaluate the whole-real-line lapse integral of the full-line kernel for $\lambda>0$. Substituting $M=N-2p/\lambda$ in the double integral of Eqs.~\eqref{eq:KN_p}--\eqref{eq:F} separates the exponent into $\lambda^{2}M^{3}/6+(\kappa-\lambda q_1)M/2$ and $4p^{3}/(3\lambda)+(\kappa-\lambda q_0)p/\lambda$~\cite{Halliwell:1988wc}. Each integral is then of Airy type,
\begin{equation}
\int_{-\infty}^{\infty}\dd t\,
e^{i(t^{3}/3+zt)}
=
2\pi\Ai(z),
\end{equation}
and the whole-real-line kernel becomes
\begin{equation}
\GR(q_1,q_0)
=
\frac{2\pi}{(2\lambda)^{1/3}}\,
\Ai\bigl(\zeta(q_1)\bigr)
\Ai\bigl(\zeta(q_0)\bigr),
\label{eq:GR_Airy}
\end{equation}
where
\begin{equation}
\zeta(q)
=
\frac{\kappa-\lambda q}{(2\lambda)^{2/3}}.
\label{eq:zeta}
\end{equation}
The functional form is that of Ref.~\cite{Halliwell:1988wc}; the coefficient $2\pi/(2\lambda)^{1/3}$ follows from the normalized measure \eqref{eq:KN_p}.
One also checks directly that $[4\partial_{q}^{2}+\lambda q-\kappa]\Ai(\zeta(q))=0$, using $\zeta'(q)=-(\lambda/4)^{1/3}$ and $\Ai''(\zeta)=\zeta\Ai(\zeta)$. This kernel is real, symmetric and positive, as required by property (i). It is the group-averaging kernel of the full-line theory and does not in general satisfy the DeWitt boundary condition.

\subsection{Constant potentials}
\label{sec:constant}
Two cosmological models considered in Ref.~\cite{Matsui:2023tkw} have a constant potential, so that the method of images applies directly. Both are
described by
\begin{equation}
S_{\mu}[x,N]=-\frac12\int_{0}^{1}\dd t\,N\left[\frac{\dot x^{2}}{4N^{2}}+\mu\right],\quad
\hC_{\mu}=2\frac{\dd^{2}}{\dd x^{2}}+\frac{\mu}{2},
\label{eq:Smu}
\end{equation}
with a constant $\mu$. The closed universe with $\lambda=0$ 
corresponds to $x=q$ and $\mu=-1$,
whereas the flat universe with $\lambda>0$ corresponds to $\mu=\lambda$
after the change of variables of Sec.~\ref{sec:flat}.
From Eq.~\eqref{eq:KN_allN},
\begin{equation}
K^{(\mu)}_{N}(x_1,x_0)=\sqrt{\frac{i}{8\pi N}}\,e^{-i(x_1-x_0)^{2}/(8N)-i\mu N/2},\quad N>0 .
\label{eq:KNmu}
\end{equation}
The positive-lapse integral is elementary. 
With $d=|x_1-x_0|$ and the regulators $e^{-\eps N-\eta/N}$, $\eps,\eta>0$, the lapse integral takes the form
\begin{equation}
\int_{0}^{\infty}\dd N\,N^{-1/2}e^{-aN-b/N}
=
\sqrt{\frac{\pi}{a}}\,e^{-2\sqrt a\sqrt b},
\end{equation}
where $a=\eps+i\mu/2$ and $b=\eta+id^{2}/8$; see Appendix~\ref{app:integral}. It follows that
\begin{equation}
\Gp^{(\mu)}(x_1,x_0)
=
\lim_{\eps,\eta\downarrow0}
\frac{\sqrt{i}}{\sqrt{8}\,\sqrt{a}}\,
e^{-2\sqrt a\sqrt b}.
\label{eq:Gplus_mu}
\end{equation}
Here the principal branches of the square roots are chosen, which are continuous throughout the convergence region $\mathrm{Re}\,a,\mathrm{Re}\,b>0$. For this integral, the one-loop saddle-point evaluation is in fact exact. Indeed, the integral can be written as $2(b/a)^{1/4}K_{1/2}(2\sqrt{ab})$ with the elementary Bessel function $K_{1/2}$. The subtleties discussed in this paper are therefore not questions of approximation but of which lapse integral is taken and how the reflection is performed.

\begin{table*}[t]
\caption{Exact DeWitt kernels of the minisuperspace models considered in the text. Here $k=\sqrt\lambda/2$, $q_{\lessgtr}=\min/\max(q_0,q_1)$, $x_{\lessgtr}=\min/\max(x_0,x_1)$, and $x=\tfrac23 q^{3/2}$.}
\label{tab:summary}
\begin{ruledtabular}
\begin{tabular}{lccc}
Model & Constraint & Positive-lapse Dirichlet Green function $\GpD$ & Whole-real-line kernel $\GRD$ \\
\hline
Closed, $\lambda=0$ & $2\partial_q^2-\tfrac12$ & $i\,e^{-q_>/2}\sinh(q_</2)$ & $0$ \\
Flat, $\lambda>0$ (variable $x$) & $2\partial_x^2+\tfrac{\lambda}{2}$ & $\dfrac{i}{\sqrt\lambda}\,e^{-ikx_>}\sin(kx_<)$ & $\dfrac{2}{\sqrt\lambda}\sin(kx_1)\sin(kx_0)$ \\
Linear potential, $\lambda>0$, any $\kappa$ & $2\partial_q^2+\tfrac{\lambda q-\kappa}{2}$ & Eq.~\eqref{eq:GpD_Airy} & Eq.~\eqref{eq:GRD_Airy} \\
\end{tabular}
\end{ruledtabular}
\end{table*}

\subsubsection{Closed universe with $\lambda=0$}
\label{sec:closed}

For $\mu=-1$ one has $\sqrt a\to e^{-i\pi/4}/\sqrt2$ and $\sqrt b\to e^{i\pi/4}d/(2\sqrt2)$, so that $2\sqrt a\sqrt b=d/2$ and $\sqrt i/(\sqrt8\sqrt a)=i/2$. Therefore,
\begin{equation}
\Gp(q_1,q_0)=\frac{i}{2}\,e^{-|q_1-q_0|/2}.
\label{eq:Gplus_closed}
\end{equation}
This is the exact positive-lapse Green function of $\hC=2\partial_{q}^{2}-\tfrac12$. Indeed, the jump $-i/2$ of $\partial_{q_1}\Gp$ at $q_1=q_0$ gives
$\hC_{q_1}\Gp=-i\delta(q_1-q_0)$. Thus, $\Gp$ satisfies the inhomogeneous Wheeler--DeWitt equation rather than the homogeneous constraint equation. It should therefore not be confused with the whole-real-line kernel. From Eqs.~\eqref{eq:Gminus} and \eqref{eq:GR_def},
\begin{equation}
\Gm=-\frac{i}{2}\,e^{-|q_1-q_0|/2},\qquad
\GR(q_1,q_0)=0 .
\label{eq:GR_closed}
\end{equation}
The vanishing of $\GR$ also follows directly from the spectral representation. In momentum space,
$C(p)=-2p^{2}-\tfrac12<0$ for all real $p$, so that
$\int\dd N\,e^{iN(2p^{2}+1/2)}
=2\pi\delta(2p^{2}+\tfrac12)$ has no support. The same conclusion holds on the half-line, where
$\mathrm{spec}\,\hCD=(-\infty,-\tfrac12]$. Hence,
\begin{equation}
\GRD(q_1,q_0)=2\pi\delta(\hCD)=0 .
\label{eq:GRD_closed}
\end{equation}

The whole-real-line DeWitt kernel therefore vanishes in this model, whereas the positive-lapse Dirichlet Green function is nonzero. Since the potential is constant, rule (a) applies. Using Eq.~\eqref{eq:image_G} with the kernel \eqref{eq:Gplus_closed}, which is valid for both orderings of the endpoints, we find for $q_0,q_1>0$
\begin{align}
\GpD(q_1,q_0)
&=\Gp(q_1,q_0)-\Gp(q_1,-q_0)\nonumber\\
&=\frac{i}{2}\Bigl[e^{-|q_1-q_0|/2}-e^{-(q_1+q_0)/2}\Bigr]\nonumber\\
&=i\,e^{-q_>/2}\sinh\frac{q_<}{2},
\label{eq:GpD_closed}
\end{align}
where $q_<=\min(q_0,q_1)$ and $q_>=\max(q_0,q_1)$. This kernel is symmetric and satisfies the Dirichlet condition in both arguments,
$\GpD(0,q_0)=\GpD(q_1,0)=0$. For $q_1\neq q_0$, it obeys
\begin{equation}
\left(2\partial_{q_1}^{2}-\frac12\right)\GpD=0,
\end{equation}
while the jump $-i/2$ of its first derivative at $q_1=q_0$ gives
$\hCD\GpD=-i\delta(q_1-q_0)$, in accordance with Eq.~\eqref{eq:D_identities}. The same result follows directly from rule (c), with
$u_D=\sinh(q/2)$ and $u_R=e^{-q/2}$.

The image term is independent of whether the initial or the final point is reflected:
\begin{equation}
\Gp(q_1,-q_0)=\Gp(-q_1,q_0)
=\frac{i}{2}e^{-(q_1+q_0)/2}.
\end{equation}
This equivalence requires using the form of the kernel that is valid for both orderings of the endpoints. In particular, for the final-point reflection one has $-q_1<q_0$, so the appropriate branch of $|q_1-q_0|$ is the one opposite to that used for the direct term when $q_1>q_0$.

The DeWitt solution of the Wheeler--DeWitt equation
$(4\partial_{q}^{2}-1)\Psi=0$ appears naturally in $\GpD$. To see this, consider a source $\chi(q_0)$ supported in
$[q_a,q_b]\subset(0,\infty)$ and define
\begin{equation}
\Psi(q_1)=\int\dd q_0\,\GpD(q_1,q_0)\chi(q_0).
\end{equation}
For $q_1<q_a$, one finds
\begin{equation}
\Psi(q_1)
=
i\sinh\frac{q_1}{2}
\int\dd q_0\,e^{-q_0/2}\chi(q_0),
\end{equation}
whereas for $q_1>q_b$,
\begin{equation}
\Psi(q_1)
=
i\,e^{-q_1/2}
\int\dd q_0\,\sinh\frac{q_0}{2}\chi(q_0).
\end{equation}
Thus, on the side of the source adjacent to the singular boundary, the Dirichlet Green function selects the DeWitt solution satisfying $\Psi(0)=0$, while beyond the source it selects the decaying solution. The growing solution $\sinh(q/2)$ is not a generalized eigenfunction of $\hCD$, since it is not polynomially bounded. This provides the spectral reason why it cannot contribute to a group-averaged kernel.

\subsubsection{Flat universe with $\lambda>0$}
\label{sec:flat}

For $\kappa=0$, the action \eqref{eq:action} reads
$S=-\tfrac12\int\dd t\,[\dot q^{2}/(4N)+\lambda Nq]$.
Setting $N=\frac{\Nt}{q}$ and $x=\tfrac23q^{3/2}$~\cite{Matsui:2023tkw},
one has $q\dot q^{2}=\dot x^{2}$, and the action becomes
\begin{equation}
S[x,\Nt]=-\frac12\int_{0}^{1}\dd t
\left[\frac{\dot x^{2}}{4\Nt}+\lambda\Nt\right].
\label{eq:flat_action}
\end{equation}
Thus, the potential is constant in the variable $x$.
The kernels computed below refer to the quantization of
Eq.~\eqref{eq:flat_action} on $L^{2}((0,\infty),\dd x)$,
with the constraint $\hC_x=2\partial_x^{2}+\lambda/2$
and the DeWitt condition $\Psi(0)=0$.
This theory differs from the quantization \eqref{eq:WDW}
in the variable $q$: the lapse rescaling by $1/q$ and
the nonlinear change of variable modify the operator ordering
and the measure. These quantum effects do not affect the
leading semiclassical phase but are retained in the exact
kernels~\cite{Hawking:1985bk,Kontoleon:1998pw}.
We return to the $q$-representation in Sec.~\ref{sec:linear}.

With $\mu=\lambda$ in Eq.~\eqref{eq:Gplus_mu},
$\sqrt a\to e^{i\pi/4}\sqrt{\lambda/2}$ and
$\sqrt b\to e^{i\pi/4}d/(2\sqrt2)$.
Hence $2\sqrt a\sqrt b=ikd$ and
$\sqrt i/(\sqrt8\sqrt a)=1/(2\sqrt\lambda)$,
where $k=\sqrt\lambda/2$.
The unrestricted full-line kernels are therefore
\begin{equation}
\Gp(x_1,x_0)=\frac{e^{-ik|x_1-x_0|}}{2\sqrt\lambda},
\
\GR(x_1,x_0)=\frac{\cos k(x_1-x_0)}{\sqrt\lambda}.
\label{eq:Gplus_flat}
\end{equation}
The positive-lapse kernel $\Gp$ is the outgoing Green function
of $\hC_x$ and satisfies
$(2\partial_{x_1}^{2}+\lambda/2)\Gp=-i\delta(x_1-x_0)$.
The whole-real-line kernel $\GR$ represents $2\pi\delta(\hC_x)$.
Indeed, the plane waves $e^{ipx}$ are eigenfunctions of $\hC_x$
with eigenvalue $\lambda/2-2p^{2}$, so that
\begin{align}
\GR(x_1,x_0)
&=
\int_{-\infty}^{\infty}\dd p\,e^{ip(x_1-x_0)}
\delta\left(\frac{\lambda}{2}-2p^{2}\right) \notag \\ 
&=
\frac{\cos k(x_1-x_0)}{2k},
\end{align}
in agreement with Eq.~\eqref{eq:Gplus_flat}.

Applying rule (a) to the kernels \eqref{eq:Gplus_flat},
which are valid for both orderings of the endpoints,
gives for $x_0,x_1>0$
\begin{align}
\GpD(x_1,x_0)
&=\frac{1}{2\sqrt\lambda}
\Bigl[e^{-ik|x_1-x_0|}-e^{-ik(x_1+x_0)}\Bigr]\nonumber\\
&=\frac{i}{\sqrt\lambda}\,e^{-ikx_>}\sin(kx_<),
\label{eq:GpD_flat}
\end{align}
and
\begin{align}
\GRD(x_1,x_0)
&=\frac{1}{\sqrt\lambda}
\Bigl[\cos k(x_1-x_0)-\cos k(x_1+x_0)\Bigr]\nonumber\\
&=\frac{2}{\sqrt\lambda}\sin(kx_1)\sin(kx_0),
\label{eq:GRD_flat}
\end{align}
where $x_<=\min(x_0,x_1)$ and $x_>=\max(x_0,x_1)$.
Both kernels vanish when either endpoint is at $x=0$.
The whole-real-line kernel $\GRD$ satisfies the homogeneous
Wheeler--DeWitt equation
$(2\partial_x^{2}+\lambda/2)\GRD=0$ in each argument.
The positive-lapse kernel $\GpD$ instead satisfies the
inhomogeneous equation with source $-i\delta(x_1-x_0)$,
which arises from the discontinuity of its first derivative
at $x_1=x_0$.

Equation~\eqref{eq:GRD_flat} can also be checked using
the normalized Dirichlet modes
$\phi_p(x)=\sqrt{2/\pi}\sin(px)$, $p>0$.
In the kernel of $2\pi\delta(\hCD)$, only the mode
$p=k=\sqrt\lambda/2$ contributes, reproducing both
the functional form and the normalization.
For $\mu<0$, however, the constraint eigenvalue
$C(p)=\mu/2-2p^{2}$ never vanishes for real $p$,
so the whole-real-line kernel is zero,
in agreement with Eq.~\eqref{eq:GRD_closed}. 
The functional form of Eq.~\eqref{eq:GRD_flat} agrees with
Eq.~(12) of Ref.~\cite{Jia:2024comment}.
The overall coefficient is fixed here by the operator identity
$\GRD=2\pi\delta(\hCD)$.
The kernel is real and defines a positive semidefinite
quadratic form, as required by property (i).

For a smooth auxiliary state $\chi$ of compact support,
the kernel \eqref{eq:GRD_flat} gives
\begin{equation}
\Psi(x_1)
=
\int_{0}^{\infty}\dd x_0\,
\GRD(x_1,x_0)\chi(x_0)
\propto\sin(kx_1).
\label{eq:Psi_flat}
\end{equation}
The resulting standing wave satisfies the DeWitt condition
and carries no current.
By contrast, the positive-lapse kernel \eqref{eq:GpD_flat}
gives a wave proportional to $e^{-ikx_1}$ above the support
of $\chi$. Since $\dot x=-4\Nt p_x$ and $p_x=-k$ for this wave,
it describes an expanding universe for $\Nt>0$.

The nodes of the wave function are located at $kx_n=n\pi$.
Since $x=\tfrac23q^{3/2}=\tfrac23a^{3}$,
this gives $a_n^{3}=3\pi n/\sqrt\lambda$.
Writing $\lambda=\sigma^{2}H_{\Lambda}^{2}$ with
$H_{\Lambda}^{2}=\Lambda/3$ and introducing the physical
scale factor $L=\sigma a$, one obtains
\begin{equation}
L_n^{3}=\frac{2nG}{H_{\Lambda}},
\qquad
L_n=\left(\frac{2n\,G}{H_{\Lambda}}\right)^{1/3}
=\bigl(2n\,\ell_{\rm P}^{2}H_{\Lambda}^{-1}\bigr)^{1/3}.
\label{eq:nodes}
\end{equation}
The node scale is therefore set by a weighted geometric mean
of the Planck length and the Hubble radius.
Because the standing wave is periodic in $x\propto a^{3}$
rather than in $a$, its nodes are not evenly spaced in the
physical size of the universe and do not mark the Hubble
radius $H_{\Lambda}^{-1}$.
For $H_{\Lambda}\ell_{\rm P}\ll1$, the first node lies
deep inside the Hubble radius, at
$L_1\sim(\ell_{\rm P}^{2}H_{\Lambda}^{-1})^{1/3}$.
The physical interpretation of this scale should nevertheless
be treated with caution within the minisuperspace approximation.

Table~\ref{tab:summary} collects the exact kernels obtained
so far, together with those of the linear potential derived
in Sec.~\ref{sec:linear}.

\subsection{Linear potential: Airy kernels}
\label{sec:linear}

We now return to the $q$-representation \eqref{eq:WDW},
with $\lambda>0$ and arbitrary $\kappa$.
The potential $\lambda q-\kappa$ is not reflection symmetric,
so the method of images cannot be applied directly to the
known full-line kernel.
The Wheeler--DeWitt equation has two independent solutions,
$\Ai(\zeta(q))$ and $\Bi(\zeta(q))$, with $\zeta$ defined in
Eq.~\eqref{eq:zeta}.
For $(\hC-i\eps)^{-1}$, the argument is shifted to
$\zeta_{\eps}=\zeta+2i\eps/(2\lambda)^{2/3}$.
For $\eps>0$, $\Ai(\zeta_{\eps})$ decays as $q\to-\infty$,
while $\Ai(\zeta_{\eps})+i\Bi(\zeta_{\eps})$ decays
as $q\to+\infty$.
In the limit $\eps\downarrow0$, the latter becomes
\begin{equation}
\begin{aligned}
F(\zeta)&\equiv\Ai(\zeta)+i\Bi(\zeta)\\
&\sim
\frac{i\,e^{-i[\frac23(-\zeta)^{3/2}+\frac{\pi}{4}]}}
{\sqrt\pi\,(-\zeta)^{1/4}},
\qquad \zeta\to-\infty.
\end{aligned}
\label{eq:Fasym}
\end{equation}
Using rule (c), together with $W_{\zeta}[\Ai,\Bi]=1/\pi$
and $\zeta'(q)=-(\lambda/4)^{1/3}$, gives
\begin{equation}
\Gp(q_1,q_0)
=\frac{\pi}{(2\lambda)^{1/3}}\;
\Ai\bigl(\zeta(q_<)\bigr)F\bigl(\zeta(q_>)\bigr).
\label{eq:Gplus_Airy}
\end{equation}
Twice its real part reproduces the whole-real-line kernel
\eqref{eq:GR_Airy}, confirming the normalization.
Equation~\eqref{eq:Gplus_Airy} also agrees with Eq.~(18)
of Ref.~\cite{Jia:2024comment} after applying the
Airy connection formulas.

Applying the rank-one formula \eqref{eq:rankone}
at finite $\eps>0$ and then taking $\eps\downarrow0$
gives, for $q_0,q_1>0$,
\begin{equation}
\GpD(q_1,q_0)
=-\frac{i\pi}{(2\lambda)^{1/3}}\;
\frac{F\bigl(\zeta(q_>)\bigr)}{F(c_0)}\;u_D(q_<),
\label{eq:GpD_Airy}
\end{equation}
where $c_0\equiv\zeta(0)=\kappa/(2\lambda)^{2/3}$ and
\begin{equation}
u_D(q)\equiv
\Ai(c_0)\Bi\bigl(\zeta(q)\bigr)
-\Bi(c_0)\Ai\bigl(\zeta(q)\bigr).
\label{eq:uD}
\end{equation}
The function $u_D$ is the DeWitt solution, satisfying
$\hC u_D=0$ and $u_D(0)=0$.
The kernel vanishes when either endpoint is at $q=0$
and satisfies $\hC_{q_1}\GpD=-i\delta(q_1-q_0)$.
For $\kappa=1$, it reduces to Eq.~\eqref{eq:GpD_closed}
as $\lambda\to0$.

Taking $\GRD=2\,\mathrm{Re}\,\GpD$, we obtain
\begin{equation}
\GRD(q_1,q_0)
=\frac{2\pi}{(2\lambda)^{1/3}}\;
\frac{u_D(q_1)\,u_D(q_0)}
{\Ai(c_0)^{2}+\Bi(c_0)^{2}}.
\label{eq:GRD_Airy}
\end{equation}
This is the kernel of $2\pi\delta(\hCD)$.
It is real, symmetric, and factorized, and defines a
positive semidefinite quadratic form.

For $\kappa=0$, Eq.~\eqref{eq:GRD_Airy} gives the flat-universe
DeWitt kernel in the $q$-representation, with $c_0=0$.
At large $q$, the Airy functions have the leading phase
$\tfrac23(-\zeta)^{3/2}=kx$, where
$x=\tfrac23q^{3/2}$ and $k=\sqrt\lambda/2$.
Thus, the DeWitt solutions in the $q$- and $x$-quantizations
share the same leading WKB phase, but differ in the
$q^{-1/4}$ prefactor and a constant phase shift,
and hence in the positions of their nodes.

Finally, applying the method of images to the even extension
$\lambda|q|-\kappa$ yields the same Dirichlet Green function
\eqref{eq:GpD_Airy}.
The two constructions agree because the boundary condition
and the $i0$ prescription fix the Green function uniquely.
For a reflection-symmetric full-line operator, the rank-one
term reduces to the image term, as shown in
Appendix~\ref{app:rankone}.

\section{Discussion}
\label{sec:discussion}
We have constructed the DeWitt kernels of minisuperspace quantum cosmology exactly, for the closed universe with $\Lambda=0$, for the flat universe with $\Lambda>0$, and for the linear potential of the closed and open de Sitter models. In each case the kernels vanish at both endpoints and 
are fixed, including their normalization, by the operator
definitions \eqref{eq:GDW_operator}.
Since the kernels have been evaluated exactly, 
what determines
the answer is not the accuracy of an approximation but which
lapse integral is taken, how the two half-lines are combined,
and how the reflection is performed. The two-sided requirement \eqref{eq:twosided} and the need to evaluate image terms with kernels valid for all orderings of the endpoints, emphasized in Ref.~\cite{Jia:2024comment}, are borne out by the construction, as is the rank-one formula \eqref{eq:rankone}, for which we have supplied a derivation and an Airy-function evaluation. We differ from Ref.~\cite{Jia:2024comment} on one point: the whole-real-line kernel of the closed $\lambda=0$ universe vanishes rather than being $-2\sinh(q_0/2)\sinh(q_1/2)$, because zero lies outside the spectrum of the constraint; the DeWitt propagator of that model is the positive-lapse Green function \eqref{eq:GpD_closed}.

The two kernels answer different questions. The positive-lapse $\GpD$ is the causal amplitude in the sense of Teitelboim~\cite{Teitelboim:1981ua,Teitelboim:1983fh} and selects the expanding branch beyond the source, Eq.~\eqref{eq:GpD_flat}, while $\GRD$ is a real standing wave with vanishing current, the natural object of refined algebraic quantization~\cite{Ashtekar:1995zh,Marolf:2000iq}; the DeWitt condition can be imposed on either. Neither becomes a wave function until it is paired with an auxiliary state $\chi$, as in Eq.~\eqref{eq:Psi_flat}, and with more degrees of freedom that pairing becomes a genuine choice of initial conditions, while the rank-one formula is replaced by an integral operator over the boundary hypersurface. We also note that the Abel regulator $e^{-\eps N}$ used here fixes the boundary value of the resolvent, whereas a Lefschetz-thimble deformation makes an oscillatory integral convergent only once the original contour and its endpoints have been specified~\cite{Witten:2010cx,Feldbrugge:2017kzv,Halliwell:1988ik,Honda:2024aro}; for quadratic actions the exact evaluation makes the deformation unnecessary.

The exact kernels also sharpen the limitations of the DeWitt condition. Its vanishing wave function at $q=0$ does not by itself imply finite expectation values of curvature invariants~\cite{Lund:1973zz,Gotay:1980xk,Gotay:1983kvm}.
The flat-universe example also shows that different quantizations
can yield different node structures despite sharing the same
leading WKB phase. Whether the construction survives the inclusion of inhomogeneous perturbations, where the DeWitt wave function of general relativity is known to be perturbatively ill-behaved~\cite{Matsui:2021yte,Martens:2022dtd}, remains to be investigated.

\begin{acknowledgments}
The author is grateful to Ding Jia for his Comment~\cite{Jia:2024comment}, which prompted this work. This work was supported by JSPS KAKENHI Grant No.~JP23K13100.
\end{acknowledgments}

\appendix

\section{The lapse integral}
\label{app:integral}
For $\mathrm{Re}\,a>0$ and $\mathrm{Re}\,b>0$ consider
\begin{equation}
I(a,b)=\int_{0}^{\infty}\dd N\,N^{-1/2}e^{-aN-b/N}=2J(a,b),
\end{equation}
\begin{equation}
J(a,b)=\int_{0}^{\infty}\dd y\,e^{-ay^{2}-b/y^{2}},
\end{equation}
where $N=y^{2}$. For real positive $a,b$, differentiating under the integral sign and substituting $y=\sqrt{b/a}/u$ gives
\begin{equation}
\frac{\partial J}{\partial b}=-\int_{0}^{\infty}\frac{\dd y}{y^{2}}\,e^{-ay^{2}-b/y^{2}}=-\sqrt{\frac{a}{b}}\;J,
\end{equation}
so that $J(a,b)=J(a,0)e^{-2\sqrt{ab}}$ with $J(a,0)=\sqrt\pi/(2\sqrt a)$, and
\begin{equation}
I(a,b)=\sqrt{\frac{\pi}{a}}\;e^{-2\sqrt a\sqrt b}.
\label{eq:Iab}
\end{equation}
Both sides are analytic in $a$ and $b$ on the right half-planes, so Eq.~\eqref{eq:Iab} holds there with the principal branches of the square roots. Equivalently, $I(a,b)=2(b/a)^{1/4}K_{1/2}(2\sqrt{ab})$ with $K_{1/2}(z)=\sqrt{\pi/(2z)}\,e^{-z}$, which also shows that the one-loop saddle-point approximation of $I(a,b)$ is exact. The positive-lapse integrals of the constant-potential models are $\sqrt{i/(8\pi)}\,I(a,b)$ with $a=\eps+i\mu/2$ and $b=\eta+i(x_1-x_0)^{2}/8$; the limits $\eps,\eta\downarrow0$ are taken after the integration, along paths on which the principal branches are continuous, which gives $\sqrt a\to e^{-i\pi/4}\sqrt{|\mu|/2}$ for $\mu<0$ and $\sqrt a\to e^{i\pi/4}\sqrt{\mu/2}$ for $\mu>0$, and $\sqrt b\to e^{i\pi/4}|x_1-x_0|/(2\sqrt2)$.

\section{Rank-one formula from an infinite barrier}
\label{app:rankone}
Let $\hC$ be the full-line constraint operator and let $R(z)=(\hC-z)^{-1}$ be its resolvent with kernel $G(q_1,q_0;z)$. Add a $\delta$-function barrier at the origin, $\hC_{\gamma}=\hC-\gamma\,|0\rangle\langle0|$, which corresponds to the potential $V\to V+\gamma\delta(q)$ in the ``Hamiltonian'' $-\hC$~\cite{Albeverio:2005}. The second resolvent identity $R_{\gamma}-R=R(\hC-\hC_{\gamma})R_{\gamma}$ reads, in kernel form,
\begin{equation}
G_{\gamma}(q_1,q_0)=G(q_1,q_0)+\gamma\,G(q_1,0)\,G_{\gamma}(0,q_0).
\end{equation}
Setting $q_1=0$ and solving for $G_{\gamma}(0,q_0)$ gives
\begin{align}
G_{\gamma}(q_1,q_0)&=G(q_1,q_0)+\frac{\gamma\,G(q_1,0)\,G(0,q_0)}{1-\gamma\,G(0,0)}\nonumber\\
&\xrightarrow{\;\gamma\to\infty\;}\;
G(q_1,q_0)-\frac{G(q_1,0)\,G(0,q_0)}{G(0,0)},
\label{eq:rankone_app}
\end{align}
provided $G(0,0)\neq0$. The limiting kernel vanishes when either argument is zero, and it is the resolvent of the operator whose domain is restricted by $\Psi(0)=0$ on both sides of the barrier, i.e.\ the direct sum of the Dirichlet operators on $(-\infty,0)$ and $(0,\infty)$; restricted to $q_0,q_1>0$ it is the resolvent of $\hCD$. Since $\Gp=-iR(i0)$, multiplying Eq.~\eqref{eq:rankone_app} by $-i$ gives Eq.~\eqref{eq:rankone}. The derivation uses only the resolvent identity; it does not apply to $e^{-iN\hC}$, and the fixed-lapse Dirichlet kernel is \emph{not} $K_{N}-K_{N}(q_1,0)K_{N}(0,q_0)/K_{N}(0,0)$.

If $\hC$ is reflection symmetric, the rank-one term reduces to the image term. In that case $G(q_1,q_0)=A\,u(q_<)\,v(q_>)$, where $u$ is the solution selected at $q\to-\infty$ and $v$ the one selected at $q\to+\infty$ by the $i0$ prescription, and parity invariance of the resolvent implies $v(q)=u(-q)$. For $q_0,q_1>0$ one then has $G(q_1,0)G(0,q_0)/G(0,0)=A\,u(0)\,v(q_1)v(q_0)/v(0)=A\,v(q_0)v(q_1)=A\,u(-q_0)v(q_1)=G(q_1,-q_0)$, using $u(0)=v(0)$. Thus Eq.~\eqref{eq:rankone} and the image formula $G(q_1,q_0)-G(q_1,-q_0)$ coincide whenever the latter is applicable, as verified explicitly by Eqs.~\eqref{eq:Gplus_closed}, \eqref{eq:GpD_closed}, \eqref{eq:Gplus_flat} and \eqref{eq:GpD_flat}.

\bibliography{dewitt_kernels}
\bibliographystyle{utphys}

\end{document}